\documentclass{jpp}
\usepackage{graphicx}
\usepackage{epstopdf, epsfig}
\usepackage[bookmarks=false]{hyperref}

\usepackage{xcolor}
\usepackage{bm}
\usepackage{amsmath}
\usepackage{natbib}
\usepackage{color}

\newcommand{\dd}{\mathrm d}
\graphicspath{{./figures/}}
\usepackage{physics}
\usepackage{amsmath}
\usepackage{enumitem}
\newcommand{\subscript}[2]{$#1 _ #2$}
\newcommand\ut{\mathsfbi{U}}
\newcommand\iit{\mathsfbi{I}}

\newcommand{\vpp}{v_{\perp}}
\newcommand{\vp}{v_{\parallel}}
\newcommand{\pp}{_{\perp}}
\newcommand{\pa}{_{\parallel}}

\newcommand{\pder}[2][]{\frac{\partial#1}{\partial#2}}

\graphicspath{{./figures/}}

\newcommand{\eik}{i\bm{k}_{\perp}\cdot\bm{\rho}}
\newcommand{\gradd}{\nabla}
\newcommand{\bes}{J_{0}(k_{\perp}\rho)}

\shorttitle{Radiative cooling in pair plasmas}
\shortauthor{D. Kennedy and P. Helander}

\title{Coulomb collisions in strongly anisotropic plasmas \\ II. Cyclotron cooling in laboratory pair plasmas }

\author{D. Kennedy\aff{1}\corresp{\email{daniel.kennedy@ipp.mpg.de}}
	\and P. Helander\aff{1}}

\affiliation{\aff{1}Max Planck Institute for Plasma Physics, D-17491 Greifswald, Germany}

\begin{document}

\maketitle

\begin{abstract}
	The behaviour of a strongly-magnetized collisional electron-positron plasma that is optically thin to cyclotron radiation is considered, and the distribution functions accessible to it on the various timescales in the system are calculated.  Particular attention is paid to the limit in which the collision time exceeds the radiation emission time, making the electron distribution function strongly anisotropic. Indeed, these are the exact conditions likely to be attained in the first laboratory electron-positron plasma experiments currently being developed, which will typically have very low densities and be confined in very strong magnetic fields. The constraint of strong-magnetization adds an additional complication in that long-range Coulomb collisions, which are usually negligible, must now be considered. A rigorous collision operator for these long-range collisions has never been written down. Nevertheless, we show that the collisional scattering can be accounted for without knowing the explicit form of this collision operator. The rate of radiation emission is calculated and it is found that the loss of energy from the plasma is proportional to the parallel collision frequency multiplied by a factor that only depends logarithmically on plasma parameters. That is, this is a self-accelerating process, meaning that the bulk of the energy will be lost in a few collision times. We show that in a simple case, that of straight field-line geometry, there are no unstable drift waves in such plasmas, despite being far from Maxwellian. 
\end{abstract}

	\section{Introduction}
	
	In a companion article \citep{KennedyHelander2020a}, hereinafter referred to as (I), it was shown that plasmas that are optically thin to cyclotron radiation relax to strongly anisotropic distributions and a theory of collisional scattering in such systems was developed. These results were general; in particular, little was assumed about the confining magnetic geometry. Nevertheless, two important conditions needed to be satisfied: 
	
		\begin{enumerate}[label=(\subscript{C}{{\arabic*}}),topsep=10pt,itemsep=1ex,partopsep=1ex,parsep=1ex]
			\item \enskip The plasma needed to be optically thin to cyclotron emission. This required the number density to be sufficiently small.
			\item \enskip  The radiation time, $\tau_{r},$ needed, at least initially, to be smaller than the collision time, $\tau_{c}.$ This required strong magnetic fields. 
		\end{enumerate}
	
	In (I), suitable candidates for systems satisfying these two conditions were proposed. In this contribution, we explore one of these suggestions: the first laboratory electron-positron plasma experiment, and develop the theory of collisional scattering in strongly-magnetized pair plasmas. 
	
	\subsection{Laboratory electron-positron plasmas}		
	
	Efforts are currently underway to create and confine the first terrestrial electron-positron plasmas in the laboratory. This is done by first accumulating positrons from a powerful source and then injecting these into a pure electron plasma confined by the dipolar magnetic field of a levitated current-carrying circular coil, so that a stationary, quasineutral electron-positron plasma is formed \citep{SunnPedersen2012}.

	It has been shown by \citet{Helander2014} that pair plasmas possess unique gyrokinetic stability properties due to the mass symmetry between the particle species. For example, drift instabilities are completely absent in straight field-line geometry, e.g., in a slab, provided that the density and temperature profiles of the two species are identical (``symmetric'' pair plasmas). The symmetry between the two species is broken if the temperature profiles of the electrons and positrons differ or there is an ion contamination. In these regimes, drift instabilities can be excited even in unsheared slab geometry \citep{Mishchenko2018}. It has also been shown that instabilities can be excited when symmetry is broken through relaxation of the quasineutrality condition \citep{Kennedy2019}. In a sheared slab, pure pair plasmas are prone to current-driven reconnecting instabilities \citep{Zocco2017}, but there are no drift waves. Note that asymmetry between the species is needed also in this case since the ambient electron flow velocity must differ from the positron one for the ambient current to be finite. 
	
	In contrast to slab geometry, a dipole magnetic field has finite curvature. In this case, the symmetry between the species is broken by curvature drifts and the plasma can be driven unstable by temperature and density gradients \citep{Helander2014}, even without ion contamination and for identical temperature profiles of the two species. This result also persists in the electromagnetic regime \citep{Helander2016}. The nonlinear stability of point dipole pair plasmas has also been addressed by \citet{Helander2017}. More recently, \citet{Mishchenko2018a} performed a detailed study of the gyrokinetic stability of pure pair plasma in both the Z-pinch and point-dipole limits. Again, it was found that such pair plasmas can be driven unstable by magnetic curvature, density and temperature gradients. These instabilities are also found in the magnetic geometry most relevant for the experiment, i.e., the magnetic field of a levitated current-carrying circular coil, as was recently demonstrated using a gyrokinetic code \citep{Kennedy2020a}.

	In spite of symmetry breaking leading to instability, there is hope that upcoming experiments themselves will actually be in a minimum energy state \citep{Helander2017} and therefore exhibit robust stability and, perhaps, little or no turbulence. If this were true, it could be the first time that a magnetically confined quasineutral plasma is free of anomalous transport \citep{WhitePaper}. Such an accomplishment would be a strong test for the predictive capabilities of magnetic confinement research and could have profound implications on our theoretical understanding. 
	
In the first electron-positron plasma experiment, the aim is to produce a plasma with density in the range 10$^{12}$m$^{-3}$ $< n <$10$^{13}$m$^{-3}$ and with a temperature $T$ between 1 and 10 eV. The Debye length $\lambda_{D} = (\epsilon_{0}T/2ne^{2})^{1/2}$ for such plasmas will therefore be on the order of a few mm and will exceed the gyroradius $\rho$ by two or three orders of magnitude provided the target magnetic field of around $B=1$ T is attained. That is, we are declaring an interest in plasmas satisfying the strongly-magnetized ordering $\rho \ll \lambda_{D}.$ In fusion plasmas, the Debye length is usually comparable to the gyroradius of the electrons but much smaller than that of the ions.   

Importantly, the very low densities, necessitated by the difficulty in sourcing and trapping large numbers of positrons, render the plasma optically thin to cyclotron radiation. The strong magnetization also means the power radiated will be relatively large. In particular, the collision time $\tau_{c}$ will tend to be much longer, at least initially, than the radiation time $\tau_{r},$ and laboratory pair plasmas will be able to dissipate large amounts of energy before collisions come into play. 

The aim of this paper is to discuss how the presence of radiative cooling will affect the equilibrium and stability of magnetically confined electron-positron plasmas. In Section \ref{sec:radiative_cooling_in_electron_positron_plasmas}, we begin by presenting a brief introduction to cyclotron cooling and introducing the relevant dynamical timescales for laboratory pair plasmas. In Section \ref{sec:3}, we begin to develop the kinetic theory of such plasmas, taking the aforementioned radiative processes into account. Section \ref{sec:4} is dedicated to a discussion of the long-range Coulomb collisions which are important in strongly magnetized plasmas, but are usually neglected. Various properties of these collisions are discussed and we prove several results about how these new collisions are included in the kinetic equation. The collisional regime is explored in Section \ref{sec:5}, before being further developed in Section \ref{sec:6} where we attempt to glean some insight on the distribution function during the collisional regime. In Section \ref{sec:7}, we repeat one of the stability calculations of \citet{Helander2014} in the strongly anisotropic limit, giving a hint of how our results impact stability properties. 

\section{Radiative cooling in electron-positron plasmas}
\label{sec:radiative_cooling_in_electron_positron_plasmas}
It was shown in (I) that a plasma that is optically thin to cyclotron emission will radiate its perpendicular kinetic energy on a timescale given by the radiation time:
\begin{equation}
\frac{\dd w_{\perp}}{\dd t} = - \frac{e^{4}B^{2}}{3\pi\epsilon_{0}(mc)^{3}} w_{\perp}, \quad \tau_{r} = \frac{3\pi\epsilon_{0}(mc)^{3}}{e^{4}B^{2}} = \frac{3\pi \rho_{e}}{\Omega_{e}^{2}c},
\label{cooling_equation}
\end{equation}
with $\Omega_{e} = eB/mc$ electron cyclotron frequency  and $\rho_{e} = v_{\text{th}e}/\Omega_{e}$ the electron gyroradius.

\subsection{Timescales in laboratory pair plasmas}

In the aforementioned laboratory efforts to create and confine the first terrestrial electron-positron plasma, high field devices with magnetic fields of order 1 T are being investigated as possible candidates for confinement. The radiative cooling time is inversely proportional to the square of the magnetic field strength and hence devices with larger magnetic fields will be able to dissipate perpendicular energy more quickly. As such, in systems of interest, the radiative cooling time can be relatively short compared to the other timescales in the system.

It is clear that collisions can mediate this cooling process due to the scattering of the velocity vector. A simple estimate of the collision time is given by 
\begin{equation}
\tau_{c} =  \frac{3}{4\sqrt{\pi}nv_{\text{th}}b_{\text{min}}^{2}\ln \Lambda},
\end{equation}
where $b_{\text{min}} = {e^{2}}/{2\pi \epsilon_{0}mv_{\text{th}}^{2}}$ is the classical distance of closest approach, $v_{\text{th}} = \sqrt{2T/m}$ is the thermal velocity and $\ln \Lambda$ is the Coulomb logarithm. The factor $3/4\sqrt{\pi}$ has been inserted to make $\tau_{c}$ equal to the conventional definition of the electron collision time by \citet{Braginskii1965}. From this estimate, and the target parameters given in the text, one can see that 
\begin{equation}
\epsilon = \frac{\tau_{r}}{\tau_{c}} = \frac{nm \ln \Lambda}{\sqrt{\pi} \epsilon_{0}B^{2}} \left(\frac{c}{v_{\text{th}}}\right)^{3} \ll 1.
\end{equation}
In the analysis below, this parameter will surface repeatedly, and will usually turn up inside a logarithm. Our results will hold to logarithmic accuracy in epsilon, whose exact definition is thus only important within a multiplicative factor of order unity. Such factors will therefore usually be ignored. An equivalent definition (again up to an order unity factor) of our expansion parameter is given by 
\begin{equation}
\epsilon = \frac{\ln \Lambda}{\sigma_{M}}\left(\frac{c}{v_{\text{th}}}\right)^{3}, 
\end{equation}
where $\sigma_{M} = B^{2}/nmc^{2}\mu_{0}.$ This quantity is the ratio of the magnetic energy to the rest mass energy and is sometimes referred to as the magnetization in the astrophysical literature (e.g., \citet{SironiSpitkovsky}). From this form, one sees immediately the constraints which must be imposed on the magnetization and temperature in order to satisfy the ordering requirements.
 
A caveat here is that, of course, the collision time itself will decrease as the plasma cools and hence this assumption will be violated after sufficient time has elapsed.

\subsection{Scattering in strongly anisotropic plasmas}

We have now seen that the first electron-positron plasma experiment aims to operate in a parameter regime where the conditions (\subscript{C}{1}) and (\subscript{C}{2}) are well satisfied. As a result, much of theory developed in (I) can be directly applied to such plasmas.

Nevertheless, an additional complication arises, namely that the plasma is strongly magnetized, $\rho \ll \lambda_{D}.$ As a result, an interesting type of collision comes into the picture and further theory must be developed.  
\section{Radiative cooling in collisional plasmas}
\label{sec:3}
In Section \ref{sec:radiative_cooling_in_electron_positron_plasmas}, we used simple estimates to show that, to leading order, the plasma is effectively collisionless. However, this is an assumption which of course will not be true indefinitely. In particular, the collision frequency, $\nu_{c}$ scales with $T^{-3/2}$ and this means that as the plasma loses thermal energy through radiative cooling, the collision frequency will increase.

 After sufficient time has elapsed, the collision time will become comparable to the radiation time and collisional effects will become important. That is, eventually the assumption $\tau_{r}\nu_{c} \ll 1$ will be violated. On even longer timescales, the collision time will be shorter than the radiation time. 

We wish to gain an understanding of how the plasma energy evolves as a function of time in these regimes. To this end, we introduce the total, perpendicular and parallel thermal energy of the plasma through the appropriate moments of the total (electron + positron)  distribution function
\begin{equation}
W = \int \frac{mv^{2}}{2} f \, \dd^{3}\bm{v}, \quad W_{\perp} = \int \frac{mv_{\perp}^{2}}{2} f \, \dd^{3}\bm{v}, \quad W_{\parallel} = \int \frac{mv_{\parallel}^{2}}{2} f \, \dd^{3}\bm{v}.
\label{energy_moments}
\end{equation}

 It is clear that this evolution of these quantities can be partitioned into three separate regimes based on the relative sizes of the collision time $\tau_{c}$ and the radiation time $\tau_{r}.$ 

\begin{enumerate}[label=(\Roman*.),topsep=10pt,itemsep=1ex,partopsep=1ex,parsep=1ex]
	\item \enskip Initially, $0 < \tau_{r} < t < \tau_{c}(0),$ the initial collision time. In this regime the behaviour is described as in (I), Section 2 therein, and we recover the result that the perpendicular energy will decay exponentially on the radiation timescale, whilst the parallel kinetic energy will remain constant. 
	\begin{equation}
	W_{\perp} = W_{\perp0} \exp (-t/\tau_{r}), \quad W_{\parallel} = \text{const.}
	\end{equation}
	\item \enskip  After some time, $t \sim \tau_{c}(t) $ and collisional scattering becomes important. The remainder of this paper is devoted to studying this process. 
	\item \enskip Eventually, $t \gg \tau_{r}, \tau_{c}(t).$ In this regime, the distribution function is isotropic (Maxwellian), and any remaining parallel kinetic energy will be converted to perpendicular kinetic energy via collisional scattering and then radiated. 
	\begin{equation}
	W \propto \exp (-t/\tau_{r}).
	\end{equation}
\end{enumerate}

\section{Collisions in strongly-magnetized pair plasmas} 
\label{sec:4}
Guided by our experience in (I), we write down the collisional kinetic equation in the varying magnetic field case 
\begin{equation}
\pder[f]{t} + v_{\parallel}\pder[f]{l} - \pder{\mu}\left(\frac{\mu}{\tau_{r}}f\right) - \frac{\mu}{m}\nabla_{\parallel}B \pder[f]{v_{\parallel}} = C[f,f],
\end{equation}
where $l$ parametrises the length of a magnetic field line, and we retain the assumption that the parallel electric field is negligible. We have also introduced $C[f,f]$ as the appropriate collision operator. It is, of course, pertinent to discuss the collisions which occur in such exotic plasmas. 

Great care must be taken as we recall that our plasmas are in an unusually strongly magnetized regime, $\rho \ll \lambda_{D}.$ As such, the standard collision operators cannot be used immediately as these invoke the opposite ordering and therefore do not ``see'' the gyromotion of the colliding particles. When the collisions are described by a Debye shielded interaction, particles can be separated by a distance even as large as $\lambda_{D}$ and still exchange momentum and energy. 

In the limit $\rho \ll \lambda_{D},$ the relevant collision operator is non-standard and must include the effects of helical trajectories. Such collisions are prevalent in highly magnetized non-neutral plasma experiments and have attracted some attention in the literature, being invoked to explain enhanced heat and particle transport in Penning-Malmberg trap experiments (e.g., \citet{Dubin1988}, \citet{Dubin1998} and \citet{ONeil1985}). However, a rigorous collision operator has never been written down. The reason for this is the presence of a novel effect, namely that there can be multiple binary collisions between the same pair of particles, rendering the usual techniques fruitless \citep{Dubin1997}. Earlier work, starting from that of \citet{IchimaruRosenbluth}, ignored this circumstance. 

However, all hope is not lost. During such long range collisions, particles execute many gyrations around the magnetic field lines. As a result of this, it follows that the magnetic moment is an adiabatic invariant \citep{Dubin1997b}:
\begin{equation}
\rho \ll \lambda _{D} \implies \mu = \frac{mv_{\perp}^{2}}{2B} = \text{constant}. \end{equation}
As such, the perpendicular kinetic energy will be conserved and these relatively long-range collisions result in negligible velocity scattering. The physical reasoning for this is as follows. When guiding centres on different field lines interact, the long-range Coulomb force causes an exchange in momentum. The force transverse to the magnetic field does almost no work, since, at leading order, the guiding centres are constrained to follow the magnetic field lines. However, the force parallel to the magnetic field can do work, causing a transfer of energy between the guiding centres. 

It is in light of this remarkable feature of the long-range collisions that we are able to make progress without knowing the full collision operator. 

Of course, standard short-range collisions, with impact parameter $b \ll \rho,$ i.e., those do that do not ``see'' the gyromotion, are still present and in fact it is these collisions which will be responsible for velocity scattering. We may therefore decompose the collision operator as
\begin{equation}
C[f,f] = C_{1}(f) + C_{2}(f)
\end{equation}
where $C_{1}(f)$ is our unknown collision operator describing long-range collisions which preserve $\vpp$ and $\mu,$ and $C_{2}(f)$ is the standard Landau operator describing short-range collisions \citep{LDLandau1936}. Each of these collision operators has an associated collision time, $\tau_{1}$ and $\tau_{2},$ respectively.  

We will make the assumption that long-range collisions are much more frequent than short-range collisions. Our aim is to prove three properties about the unknown collision operator $C_{1}$ and how it acts in tandem with the Landau collision operator $C_{2}.$ 

		\begin{enumerate}[label=(\subscript{P}{{\arabic*}}),topsep=10pt,itemsep=1ex,partopsep=1ex,parsep=1ex]
	\item \enskip The collision operator $C_{1}$ Maxwellianises $\vp$ for each $\vpp.$ 
	\item \enskip  Although $\tau_{1} \ll \tau_{2}$ initially, eventually the short-range collisions become important.
	\item \enskip The above properties also hold in general magnetic geometry.
\end{enumerate}

These properties will allow us to make progress without explicitly writing down the collision operator $C_{1}.$ 

\subsection{ \texorpdfstring{$P_{1}$}{TEXT}:  \texorpdfstring{$C_{1}$}{TEXT} Maxwellianises  \texorpdfstring{$\vp$}{TEXT} for each  \texorpdfstring{$\vpp$}{TEXT}}

The second law of thermodynamics demands that the action of any collision operator must not decrease the entropy of a system, $C_{1}$ is of course no exception.

Consider the entropy functional 
\begin{equation}
S[f] = - \int f \ln f \, \dd^{3} \bm{v}.
\label{entropy_functional}
\end{equation}
Then any distribution function $f_{0}(\bm{r},\bm{v})$ say, which has been allowed to evolve under the influence of long-range Coulomb collisions, must maximise this functional subject to certain constraints. 

As with any collision operator, the total particle number and the total energy must be conserved. That is, 
\begin{gather}
\int f_{0} \, \dd^{3}\bm{v} = n(\bm{r}), \quad \text{fixed}. \label{cons1}\\
\int \frac{1}{2}mv^{2}f_{0} \, \dd^{3}\bm{v} = \frac{3n(\bm{r})T(\bm{r})}{2}, \quad \text{fixed}.
\end{gather}

We also know that the magnetic moment is conserved during long-range collisions and hence an additional constraint is 
\begin{equation}
\int f_{0} \, \dd v_{\parallel} = f_{\perp}(v_{\perp}), \quad\text{fixed}.
\label{cons3}
\end{equation}
In order to find the distribution function $f_{0}$ which maximises (\ref{entropy_functional}) subject to the constraints (\ref{cons1}) - (\ref{cons3}), we introduce sets of Lagrange multipliers $\kappa(\bm{r}), \, \lambda(\bm{r})$ and $\xi (v_{\perp})$ and seek to maximise the functional 
\begin{align}
&W[f_{0},\kappa,\lambda,\xi] = - \int f_{0} \ln f_{0} \, \dd^{3}\bm{v} + \kappa(\bm{r}) \left[ \int f_{0} \, \dd^{3}\bm{v} - n(\bm{r}) \right]  \nonumber \\ &+  \lambda(\bm{r}) \left[ \int \frac{1}{2}mv^{2}f_{0} \, \dd^{3}\bm{v} - \frac{3n(\bm{r})T(\bm{r})}{2}\right] + \int \xi(v_{\perp}) \left[ \int f_{0} \, \dd v_{\parallel} - f_{\perp}(v_{\perp}) \right] \, \dd v_{\perp}.
\end{align}
The first order variation is given by 
\begin{equation}
\delta W = \int \delta f \left(-(1+\ln f_{0}) + \kappa(\bm{r}) + \frac{1}{2}mv^{2}\lambda(\bm{r}) + \xi(v_{\perp})\right) \,\dd^{3} \bm{v}.
\end{equation}
In order to ensure that this variation vanishes, the integrand must vanish point-wise and hence we obtain 
\begin{equation}
f_{0}(\bm{r},\bm{v}) = \mathrm{e}^{-\kappa(\bm{r}) + 1} \mathrm{e}^{-\frac{1}{2}mv_{\parallel}^{2}\lambda(\bm{r})} \mathrm{e}^{-\frac{1}{2}mv_{\perp}^{2}\lambda(\bm{r})} \mathrm{e}^{-\xi(v_{\perp})} = f_{\parallel}(v_{\parallel})f_{\perp}(v_{\perp}).
\end{equation}
Indeed, this implies that the distribution function can be decomposed as 
\begin{equation}
f(v_{\parallel},v_{\perp}) = n f_{\parallel}(v_{\parallel})f_{{\perp}}(v_{\perp}), \quad f_{\parallel}(v_{\parallel}) = \left(\frac{m}{2\pi T_{\parallel}}\right)^{1/2}\exp\left( -\frac{mv_{\parallel}^{2}}{2T_{\parallel}}\right).
\end{equation} 

\subsection{ \texorpdfstring{$P_{2}$}{TEXT}: short-range collisions are still important}

In a straight field, the kinetic equation is
\begin{equation}
\pder[f]{t} - \frac{1}{\tau_{r}}\pder{\mu} (\mu f) = C_{1}(f) + C_{2}(f)
\end{equation}
where $C_{1}$ describes collisions that Maxwellianise $\vp$ for each $\vpp.$ We write 
\begin{equation}
f = g \mathrm{e}^{t/\tau_{r}}, \quad \frac{\vpp^{2}}{2} = y\mathrm{e}^{-t/\tau_{r}}
\end{equation}
and find that
\begin{equation}
\left(\pder[f]{t}\right)_{\mu} = \left(\pder[g]{t}\right)_{y}\mathrm{e}^{t/\tau_{r}},
\end{equation}
from which it then follows that
\begin{equation}
\pder[g]{t} = C_{1}(g) + C_{2}(g).
\end{equation}
If
\begin{equation}
C_{1}(f) \sim \frac{f}{\tau_{1}}, \quad C_{2}(f) \sim \frac{f}{\tau_{2}}, \quad \tau_{1}\ll \tau_{2}
\end{equation}
then $g$ will approach a parallel Maxwellian on the timescale $\tau_{1},$ that is
\begin{equation}
f(\vp,y,t)\rightarrow f_{\pp}(y,t) \sqrt{\frac{m}{2\pi T_{\parallel}(t)}} \exp \left(-\frac{m\vp^{2}}{2T_{\parallel}(t)}\right), \quad t \gg \tau_{1}
\end{equation}

The ``thermal'' (i.e., typical) velocity in the perpendicular direction is thus given by 
\begin{equation}
\vpp \sim \vp \mathrm{e}^{-t/\tau_{r}}, \quad \vp = \sqrt{\frac{2T_{\parallel}}{m}}.
\end{equation}

It can be shown (see (I)) that the operator $C_{2}$ then approaches
\begin{equation}
C_{2}(f) \rightarrow \frac{\sigma}{n} \sqrt{\frac{m}{\pi T_{\parallel}}} |\ln \epsilon| \frac{1}{\vpp}\pder{\vpp}\vpp \pder[f]{\vpp}, \quad \sigma = \frac{ne^{4}\ln \Lambda}{8\pi\epsilon_{0}^{2}m^{2}}
\end{equation}
where $\ln \Lambda$ is the Coulomb logarithm. It follows that in the new coordinate system 
\begin{equation}
C_{2}(f) \rightarrow \frac{2\sigma}{n} |\ln \epsilon| \sqrt{\frac{m}{\pi T_{\parallel}}} \pder{y} \left(y\pder[f]{y}\right) \mathrm{e}^{t/\tau_{r}} \sim \frac{\mathrm{e}^{t/\tau_{r}}}{\tau_{2}}  |\ln \epsilon | f 
\end{equation}
where $\tau_{2}$ is equal to $\tau_{c}$ evaluated at the temperature $T_{\pa}.$

Hence, it is clear that $C_{2}(f)$ is only smaller than the radiation term as long as 
\begin{equation}
\mathrm{e}^{t/\tau_{r}} |\ln \epsilon| \ll \frac{\tau_{2}}{\tau_{r}} \implies t \ll \tau_{r} \ln \left( \frac{\tau_{2}}{\tau_{r}|\ln\epsilon|} \right). 
\end{equation}
At later times, $C_{2}(f)$ cannot be neglected. The distribution will then stop contracting in the perpendicular direction. 

\subsection{ \texorpdfstring{$P_{3}$}{TEXT}: general magnetic geometry} 

In (I), it was shown that the theory of collisional scattering in strongly anisotropic plasmas could be developed for general magnetic geometry. This result also holds in strongly-magnetized plasmas. 

We were able to show in (I), albeit with different collision operators, that bounce-averaging the collisional kinetic equation with a varying magnetic field results in the equation
\begin{eqnarray}
\pder[\overline{f}]{t} - \frac{1}{\tau_{r}}\overline{\pder{\mu}(\mu f)} - \frac{\mu B_{0}}{\tau_{r}}\overline{\pder[f]{w}} = \overline{C_{1}(f)} + \overline{C_{2}(f)},
\end{eqnarray}
where the bounce average is defined as 
\begin{equation}
\overline{Q} = \frac{1}{\tau_{b}} \int Q \, \frac{\dd l}{\vp}, \quad \tau_{b} = \int \frac{\dd l}{\vp}
\end{equation}
and we are to understand that the $B^{2}$ term in Larmor's formula is replaced with its average along a field line. 

We again take $C_{1},$ denoting $\mu-$preserving collisions, to be dominant, and expand $\overline{f} = \overline{f_{0}} + \overline{f_{1}} + \cdots$ to obtain 
\begin{eqnarray}
\pder[\overline{f_{0}}]{t} - \frac{1}{\tau_{r}}\overline{\pder{\mu}(\mu f_{0})} - \frac{\mu B_{0}}{\tau_{r}}\overline{\pder[f_{0}]{w}} = \overline{C_{1}(f_{0})}. 
\end{eqnarray}
We again switch to a stretched coordinate system and write 
\begin{equation}
\overline{f_{0}} = g\mathrm{e}^{t/\tau_{r}}, \quad \mu = B_{0}^{-1}y\mathrm{e}^{-t/\tau_{r}},
\end{equation}
such that 
\begin{equation}
\overline{\left(\pder[f_{0}]{t}\right)_{\mu} - \frac{1}{\tau_{r}}\pder{\mu}(\mu f_{0})} = \mathrm{e}^{t/\tau_{r}} \left(\pder[\overline{g}]{t}\right)_{y}
\end{equation}
and we thus obtain 
\begin{equation}
\pder[\overline{g}]{t} - \frac{y\mathrm{e}^{-t/\tau_{r}}}{\tau_{r}} \pder[\overline{g}]{w} = \overline{C_{1}(g)}.
\label{null_space}
\end{equation}
Equation (\ref{null_space}) suggests that $g$ will approach a distribution function in the null space of the bounce-averaged collision operator $\overline{C_{1}}.$ However, any such function must satisfy
\begin{align}
0 &= - \int \tau_{b} \frac{4\pi}{m^{2}} \overline{C_{1}(g)} \ln g \,  \dd w \dd \mu = - \frac{4\pi}{m^{2}} \int \ln g \, \dd w \dd \mu \oint C_{1}(g) \, \frac{\dd l}{\vp} \nonumber \\ &= - \oint \frac{\dd l}{B} \int \ln g C(g) \, \dd^{3}\bm{v} \geq 0
\end{align}
with equality if, and only if, $g$ is of the form 
\begin{equation}
g = h (\mu,t) \mathrm{e}^{-w/T(\mu,t)}.
\end{equation}
Once this relaxation has occurred, on the time scale $\tau_{1},$ $f$ will be Maxwellian in the parallel direction for each $\mu,$ and the distribution will have shrunk so much in the $\mu$ direction that $C_{2}$ becomes important. 

We have now proven that properties \subscript{P}{1} and \subscript{P}{2} also hold in general magnetic geometry. We have also dealt with the general magnetic geometry case for the Landau collision operator in (I).

\subsection{Collision operator in a strongly-magnetized plasma}

The short-range collisions, which have not yet been taken into account, can be described by the usual Landau collision operator acting on the decomposed distribution functions. This leads us to a result which will be the key in unlocking this problem.

Namely, the appropriate treatment of the collision operator in a strongly-magnetized plasma is simply 
\begin{equation}
C(f) = C_{1}(f) + C_{2}(f) \simeq C_{2}\left(n f_{\parallel}(v_{\parallel})f_{\perp}(v_{\perp})\right), \quad t \gg \tau_{1} \gg \tau_{r} \label{431}\end{equation} 
with 
\begin{equation}
f_{\parallel} =  \left(\frac{m}{2\pi T_{\parallel}}\right)^{1/2} \exp \left({-\frac{mv_{\parallel}^2}{2T_{\parallel}}} \right), \quad \int f_{\perp} \, \dd^{2}v_{\perp} = 1. \label{432}
\end{equation}

 We can now turn our attention to solving the collisional kinetic equation. 

\section{Collisional scattering in pair plasmas}
\label{sec:5}
The kinetic equation is
\begin{equation}
\pder[f]{t} + v_{\parallel}\pder[f]{l} + \pder{\mu}\left(\frac{\mu}{\tau_{r}}f\right) - \frac{\mu}{m}\nabla_{\parallel}B \pder[f]{v_{\parallel}}  \simeq C_{2}(f), 
\end{equation}
where we now understand that the long-range collisions have been taken into account by invoking equation (\ref{431}).

In the interest of adopting a pedagogical approach, let us specialise to straight field line geometry, noting once more that the results should hold more generally after associating $B^{2}$ with its field line average. 

One then arrives at the equation

\begin{equation}
\pder[f]{t} - \frac{1}{\tau_{r}}\pder{\mu}(\mu f) = \frac{{\sigma}}{n} \int \nabla \cdot \left[\ut \cdot (f^{\prime} \gradd f - f \gradd^{\prime} f^{\prime})  \right] \,\dd^{3}\bm{v}^{\prime}.
\end{equation}
The gradient operators are defined by
\begin{equation}
\gradd = \pder{\bm{v}}, \quad \gradd^{\prime} = \pder{\bm{v}^{\prime}}.
\end{equation}
We have also introduced $\ut,$ the second-rank tensor
\begin{equation}
\ut(\bm{u}) = \gradd\gradd u= \frac{u^{2} \iit - \bm{u}\bm{u}}{u^{3}},
\end{equation}
where $\bm{u} = \bm{v} - \bm{v}^{\prime}$ is the difference in velocity vectors between colliding particles, and $\iit$ is the identity matrix. 

The energy moments introduced in equation (\ref{energy_moments}) satisfy the evolution equations
\begin{equation}
\frac{\dd W}{\dd t} = - \frac{W_{\perp}}{\tau_{r}}, \quad \frac{\dd W_{\perp}}{\dd t} = - \frac{W_{\perp}}{\tau_{r}}+ S_{c},
\label{evolution_energy}
\end{equation} where the scattering term $S_{c}$ is given by
\begin{equation}
S_{c} = \int \frac{mv_{\perp}^{2}}{2} C[f,f] \, \dd^{3}\bm{v} = -\frac{\sigma m}{n} \iint \bm{v}_{\perp} \cdot \left[\ut \cdot (f^{\prime} \gradd f - f \gradd^{\prime} f^{\prime})  \right] \, \dd^{3}\bm{v} \, \dd^{3}\bm{v}^{\prime}.
\end{equation}

We can now take advantage of the large anisotropy in the distribution function, which is a result of the perpendicular kinetic energy being radiated before collisions become important, by expanding the scattering operator in powers of the small parameter 
\begin{equation}
\epsilon = \frac{u_{\perp}}{u_{\parallel}} \ll 1.
\label{508}
\end{equation} 
To lowest order, this yields 
\begin{align}
S_{c} &=  -\frac{{\sigma}m}{n} \iint \bm{v}_{\perp}   \cdot  \frac{u^{2} \iit - \bm{u}\bm{u}}{u^{3}} \left( f^{\prime} \gradd f - f \gradd^{\prime} f^{\prime}  \right) \, \dd^{3}\bm{v} \, \dd^{3}\bm{v}^{\prime}, \nonumber \\ &\simeq -\frac{\sigma m}{n} \iint \frac{\bm{v}_{\perp}}{u_{\pa}} \cdot f^{\prime}\gradd f \, \dd^{3} \bm{v}\, \dd^{3} \bm{v}^{\prime},
\nonumber\\&\simeq \frac{2\sigma m}{n} \iint \frac{ff^{\prime}}{u_{\parallel}} \, \dd^{3}\bm{v} \, \dd^{3}\bm{v}^{\prime}
\label{509}
\end{align}
and one can evaluate
\begin{align}
\iint \frac{ff^{\prime}}{u_{\parallel}} \, \dd^{3}\bm{v} \, \dd^{3}\bm{v}^{\prime} &= n^{2} \left(\frac{m}{2\pi T_{\parallel}}\right) \int_{-\infty}^{\infty}\int_{-\infty}^{\infty} \exp \left(-\frac{m(v_{\parallel}^{2} + v_{\parallel}^{\prime 2})}{2T_{\parallel}}\right) \frac{\dd v_{\parallel} \, \dd v_{\parallel}^{\prime}}{| v_{\parallel} - v_{\parallel^{\prime}}|} \\ &= \frac{n^{2}}{\pi} \sqrt{\frac{m}{2T_{\parallel}}} \int_{-\infty}^{\infty} \int_{-\infty}^{\infty} \exp (-x^{2} - y^{2}) \frac{\dd x \, \dd y}{|x-y|},
\label{divergent}
\end{align}
where we have made the substitution 
\begin{equation}
x = v_{\parallel}\sqrt{\frac{m}{2T_{\parallel}}}, \quad y = v_{\parallel}^{\prime}\sqrt{\frac{m}{2T_{\parallel}}},
\end{equation}
and note that a small region around $x=y$ needs to be excluded from the integration range since the ordering (\ref{508}) does not hold.
 
Upon making a further change of variables
\begin{align}
x - y &= u\sqrt{2}, \\ x + y &= v\sqrt{2},
\end{align}
we obtain
\begin{equation}
\iint \frac{ff^{\prime}}{u} \, \dd^{3}\bm{v} \, \dd^{3}\bm{v}^{\prime} = \frac{n^{2}}{2\pi}\sqrt{\frac{m}{T_{\parallel}}} \int_{-\infty}^{\infty} \frac{\exp(-u^2)}{u} \, \dd u \int_{-\infty}^{\infty}\exp(-v^{2}) \, \dd v = \frac{n^{2}}{2}\sqrt{\frac{m}{\pi T_{\parallel}}} | \ln \epsilon |.
\label{515}
\end{equation}
Here it is pertinent to remark that we had a choice in whether to perform the integration over $\vp$ or $\vpp$ first in equation(\ref{509}). Since we know $f_{\perp},$ we could perform the $\vp$ integrals first, giving a Bessel function instead of the divergent integral (\ref{divergent}), thus avoiding the ``ad-hoc'' cut off of the latter. This procedure is displayed explicitly in the next section.

It follows from equation (\ref{515}), that provided $w_{\perp} \ll w,$ we have
\begin{equation}
S_{c} = \sigma m n \sqrt{\frac{m}{\pi T_{\parallel}}} |\ln \epsilon|  = \frac{3}{\sqrt{2}} \frac{W}{\tau_{2}} |\ln \epsilon|
\label{energy_evolution_scattering}
\end{equation}
where $W \simeq nT_{\pa}/2$ and $\tau_{2} \propto T_{\pa}^{3/2}.$ 

It then follows from equation (\ref{evolution_energy}) that
\begin{equation}
\frac{\dd W}{\dd t} = - \frac{W_{\perp}}{\tau_{r}} = - S_{c} \simeq - \frac{3}{\sqrt{2}}\frac{W}{\tau_{0}}\left(\frac{W_{0}}{W}\right)^{3/2} | \ln \epsilon |,
\label{first_order_scattering_rate}
\end{equation}
where $\tau_{0}$ denotes the collision time $\tau_{c}$ when $W = W_{0}.$ In deriving this equation, we have appealed to the properties of the two collision operators discussed in Section \ref{sec:4}. 

Equation (\ref{first_order_scattering_rate}) suggests that
\begin{equation}
\frac{W}{W_{0}} = \left(1 - \frac{9|\ln \epsilon| t}{2^{3/2}\tau_{0}}\right)^{2/3}
\end{equation} and thus $W \rightarrow 0$ after a few collision times, after which $W_{\perp} \ll W$ will no longer hold. Instead 
\begin{equation}
\frac{\dd W}{\dd t} \simeq - \frac{2W}{3\tau_{r}}
\end{equation}
and $W$ will fall exponentially.

Equation (\ref{first_order_scattering_rate}) describes a self-accelerating process and the plasma will only remain in this regime for a few collision times $\tau_{c}.$

\subsection{Radiative cooling in electron-positron plasmas and implications in the laboratory}

One can now elucidate the different cooling regimes which were introduced at the beginning of this section. 

\begin{enumerate}[label=(\Roman*.),topsep=10pt,itemsep=1ex,partopsep=1ex,parsep=1ex]
	\item \enskip Initially, $0 < \tau_{r} < t < \tau_{c}(0),$ the initial collision time.
	\begin{equation}
	W_{\perp} = W_{\perp,0} \exp (-t/\tau_{r}), \quad W_{\parallel} = \text{constant.}
	\end{equation}
	\item \enskip   Then, $t \sim \tau_{c}(t) $ and scattering will occur. Equation (\ref{first_order_scattering_rate}) will hold and hence 
	\begin{equation}
W = W_{0}\left(1 - \frac{9 | \ln \epsilon| t}{2^{3/2}\tau_{0}}\right)^{2/3}
	\end{equation} 
	This is a self-accelerating process. The power radiated by the plasma in this regime can be calculated through 
	\begin{equation}
	-\frac{\dd W}{\dd t} \simeq  \frac{3}{\sqrt{2}}\frac{W}{\tau_{0}}\left(\frac{W_{0}}{W}\right)^{3/2} | \ln \epsilon |.
	\end{equation}
	
	\item \enskip Eventually, $t \gg \tau_{r}, \tau_{c}(t).$ In this regime any remaining parallel kinetic energy will be converted to perpendicular kinetic energy via collisional scattering and then radiated
	\begin{equation}
	W \propto \exp (-t/\tau_{r}).
	\end{equation}
\end{enumerate}

Thus, it seems as though the exploitation of cyclotron cooling at high field will provide a very efficient mechanism to dissipate heat in the plasma as the radiation will simply be absorbed by the vessel walls. 

This rapid cooling will have another benefit for laboratory experiments. As stated previously, there are fairly stringent conditions on the number of positrons which one can accumulate and store in the laboratory. As such, this means that the laboratory plasmas will have extremely low densities. In order for the system to be classified as a plasma, as opposed to simply a collection of charged particles, there must be collective behaviour which places a requirement on the Debye length compared to the system size $L.$ Namely, one must ensure that
\begin{equation}
\lambda_{D} = \sqrt{\frac{\epsilon_{0}T}{2ne^{2}}} \ll L.
\end{equation} 
The conditions placed on both $L$ and $n$ ensure that meeting this requirement could prove difficult. However, cyclotron cooling opens up a new avenue through which this condition might be satisfied. Simply put, the cyclotron cooling process will lower the Debye length on the time scale $\max(\tau_{r},\tau_{c}(0))$ and convert what might (and indeed likely will) be initially a collection of electron-positron pairs, into a plasma. 

For a collisionless pair plasma, we require a condition on the plasma parameter:
\begin{equation}
\Lambda = n \lambda_{D}^{3} \gg 1,
\label{Debyesphere}
\end{equation}
i.e., that there are many particles in a Debye sphere. For a fixed density, this parameter scales with $T^{3/2}$ and so this condition will become less and less well satisfied. Inevitably, there will be some trade-off between the density and temperature, which dictates the kind of plasma we are able to explore. In the upcoming series of laboratory experiments, this will not be a serious limitation; the Debye length will not be sufficiently small as to violate (\ref{Debyesphere}).

\subsection{A note on entropy}

It is interesting to look at the rate of change of entropy in optically thin electron-positron plasmas. Defining the entropy $S$ by 
\begin{equation}
S= -\int f \ln f \, \dd^{3} \bm{r} \, \dd^{3} \bm{v},
\end{equation}
it follows immediately from the kinetic equation that 
\begin{equation}
\frac{\dd S}{\dd t} + \frac{n}{\tau_{r}} = -\int C[f,f] \ln f \, \dd^{3}\bm{r} \, \dd^{3}\bm{v} \geq 0
\end{equation}
where the final inequality follows from Boltzmann's H-theorem. 

This tells us that, on the timescale $t \sim \tau_{r} \ll \tau_{c}/|\ln\epsilon|$ the entropy of the plasma decreases at a constant rate that is proportional to the density. There is a simple physical explanation of this result, namely, that the entropy of the plasma will decrease through the loss of energy and that the loss rate will be proportional to the number of the particles in the plasma. Note also that the entropy loss rate thus remains constant as the plasma loses energy, which can be understood from the fact that the energy loss $\dd Q,$ in the thermodynamic relation 
\begin{equation}
\dd S = \frac{\dd Q}{T},
\end{equation}
is proportional to temperature for cyclotron radiation. 

\section{The perpendicular component of the distribution function}\label{sec:6}

We have already argued that, for times exceeding $\tau_{1},$
\begin{equation}
f(v_{\parallel},v_{\perp}) = n f_{\parallel}(v_{\parallel})f_{{\perp}}(v_{\perp}), \quad f_{\parallel}(v_{\parallel}) = \left(\frac{m}{2\pi T_{\parallel}}\right)^{1/2}\exp\left( -\frac{mv_{\parallel}^{2}}{2T_{\parallel}}\right),
\end{equation} 
where the long-range guiding centre collisions have driven $f_{\parallel}$ to a Maxwellian. It is instructive to ask what can be deduced about $f_{\perp}.$ 

We will begin from the kinetic equation
\begin{equation}
\pder[f]{t} - \frac{1}{\tau_{r}}\pder{\mu}(\mu f) = \frac{\sigma}{n} \int \nabla \cdot \left[\ut \cdot (f^{\prime} \gradd f - f \gradd^{\prime} f^{\prime})  \right] \,\dd^{3}\bm{v}^{\prime}.
\end{equation}

Integrating over the parallel velocity to remove the $f_{\parallel}$ terms on the left-hand-side, one obtains
\begin{equation}
\pder[f_{\perp}]{t} - \frac{1}{2\tau_{r}v_{\perp}}\pder{v_{\perp}}\left(v_{\perp}^{2}f_{\perp}\right) = C_{\perp}[f,f],
\label{perpendicular_equation}
\end{equation}
where we have now introduced the operator
\begin{equation}
C_{\perp}[f,f] = \frac{\sigma}{n^{2}} \frac{1}{v_{\perp}} \pder{v_{\perp}} \bm{v}_{\perp} \cdot\int_{-\infty}^{\infty} \dd v_{\parallel} \int f f^{\prime} \ut \cdot \left(\gradd\ln f - \gradd^{\prime} \ln f^{\prime}\right) \, \dd^{3}\bm{v}^{\prime}.
\end{equation}
Now, using the factorisation of the distribution function we have
\begin{equation}
C_{\perp} = \sigma \frac{1}{v_{\perp}} \pder{v_{\perp}} \bm{v}_{\perp} \cdot \int f_{\perp}f_{\perp}^{\prime} \dd^{2}\bm{v}_{\perp}^{\prime} \iint f_{\parallel} f_{\parallel}^{\prime} \ut \cdot \left(\gradd\ln f_{\parallel} - \gradd^{\prime} \ln f_{\parallel}^{\prime} + \gradd_{\perp} \ln f_{\perp} - \gradd_{\perp}^{\prime} \ln f_{\perp}^{\prime}\right) \, \dd \vp \dd \vp^{\prime}
\end{equation}
and we can also exploit that $f_{\parallel}$ is a Maxwellian to obtain \begin{equation}
\gradd_{\parallel} \ln f_{\parallel}  - \gradd^{\prime}_{\parallel} \ln f^{\prime}_{\parallel} = \frac{m}{T_{\parallel}}(\bm{v}_{\parallel}^{\prime} - \bm{v}_{\parallel}) = - \frac{m\bm{u}_{\parallel}}{T_{\parallel}}.
\end{equation}
We thus arrive at 
\begin{align}
C_{\perp} = \sigma \frac{1}{v_{\perp}} \pder{v_{\perp}} \bm{v}_{\perp} \cdot \int f_{\perp}f_{\perp}^{\prime} \dd^{2}\bm{v}_{\perp}^{\prime} &\iint f_{\parallel} f_{\parallel}^{\prime} \Biggl[ \frac{m}{T_{\parallel}} \left(-\frac{u^{2}  \bm{u}_{\parallel} - u_{\parallel}^{2}\bm{u}}{u^{3}}\right) \nonumber \\  &+ \ut \cdot (\gradd_{\perp} \ln f_{\perp} - \gradd_{\perp}^{\prime} \ln f_{\perp}^{\prime}) \Biggl] \, \dd v_{\parallel} \dd v_{\parallel}^{\prime},
\end{align}
and the leading order (in $\epsilon \ll 1$) term is given by
\begin{equation}
C_{\perp} \simeq  \sigma \frac{1}{v_{\perp}} \pder{v_{\perp}} v_{\perp}\pder[f_{\perp}]{v_{\perp}} \int f_{\perp}^{\prime} \, \dd^{2} \bm{v}_{\perp}^{\prime} \iint \frac{f_{\parallel} f_{\parallel}^{\prime}}{u} \, \dd v_{\parallel} \, \dd v_{\parallel}^{\prime}.
\end{equation}

We first turn our attention to 
\begin{equation}
I_{1}  := \iint \frac{f_{\parallel} f_{\parallel}^{\prime}}{u} \, \dd v_{\parallel} \, \dd v_{\parallel}^{\prime}.
\end{equation}
Upon making the change of variables 
\begin{gather}
v_{\parallel} - v_{\parallel}^{\prime} = u_{\parallel},
\\
v_{\parallel} + v_{\parallel}^{\prime} = w_{\parallel},
\end{gather}
we see that we may write
\begin{equation}
f_{\parallel}f_{\parallel}^{\prime} = \frac{m}{2\pi T_{\parallel}} \exp \left(-\frac{m}{4T_{\parallel}} \left(u_{\parallel}^{2} + w_{\parallel}^{2}\right)\right),
\end{equation}
and thus our integral becomes 
\begin{equation}
I_{1} = \frac{m}{2 \pi T_{\parallel}} \int_{-\infty}^{\infty} \frac{\exp \left(-\frac{mu_{\parallel}^{2}}{4T_{\parallel}}\right)}{\sqrt{u_{\parallel}^{2} + u_{\perp}^{2}}} \, \dd u_{\parallel} \int_{-\infty}^{\infty} \exp \left(-\frac{mw_{\parallel}^{2}}{4T_{\parallel}}\right) \, \dd w_{\parallel}.
\end{equation}
Upon evaluating the second integral term and making a further change of variables $x^{2} = mu_{\parallel}^{2}/4T_{\parallel},$ we see that \begin{equation}
I_{1} = \sqrt{\frac{m}{\pi T_{\pa} }}\int_{-\infty}^{\infty} \frac{\mathrm{e}^{-x^2}}{\sqrt{x^{2} + \frac{mu_{\perp}^{2}}{4T_{\parallel}}}}\, \dd x = \sqrt{\frac{m}{\pi T_{\pa}}} \mathrm{e}^{\frac{mu_{\perp}^{2}}{8T_{\parallel}}} K_{0} \left(\frac{mu_{\perp}^{2}}{8T_{\parallel}}\right),
\end{equation}
where $K_{0}$ is the modified Bessel function of the second kind. 

In order to calculate $C_{\perp}$ it remains to evaluate $I_{2}$ given by 
\begin{equation}
I_{2} := \sqrt{\frac{m}{\pi T_{\pa}}}\int f_{\perp}^{\prime} \mathrm{e}^{\frac{mu_{\perp}^{2}}{8T_{\parallel}}} K_{0} \left(\frac{mu_{\perp}^{2}}{8T_{\parallel}}\right) \, \dd^{2} \bm{v}_{\perp}^{\prime}.
\end{equation}
Introducing the change of variables 
\begin{equation}
x_{\perp}^{2} = \frac{mv_{\perp}^{2}}{2T_{\perp}}, \quad x_{\perp}^{\prime 2} = \frac{mv_{\perp}^{\prime 2}}{2T_{\perp}},
\end{equation}
we obtain 
\begin{equation}
\frac{mu_{\perp}^{2}}{8T_{\parallel}} = \delta (x_{\perp}^{2} + x_{\perp}^{\prime 2} - 2x_{\perp}x_{\perp}^{\prime}\cos\theta) =: \delta g(x_{\perp},x_{\perp}^{\prime},\theta),
\end{equation}
where $\theta$ is the angle between $\bm{v}_{\perp}$ and $\bm{v}_{\perp}^{\prime}$ and we have introduced another small parameter \begin{equation}
\delta = \frac{T_{\perp}}{4T_{\parallel}} \ll 1.
\end{equation}
We may now approximate 
\begin{equation}
I_{2} = \frac{2T_{\perp}}{m} \int_{0}^{\infty} f_{\perp}^{\prime} x_{\perp}^{\prime} \, \dd x_{\perp}^{\prime} \int_{0}^{2\pi} \mathrm{e}^{\delta g(x_{\perp},x_{\perp}^{\prime},\theta) } K_{0} (\delta g(x_{\perp},x_{\perp}^{\prime},\theta)) \, \dd \theta.
\end{equation}
To leading order in $\delta,$ we thus obtain, to logarithmic accuracy,
\begin{equation}
I_{2} \simeq \sqrt{\frac{m}{\pi T_{\pa}}} |\ln \delta|,
\end{equation} 
and hence
\begin{equation}
C_{\perp} = \sigma \sqrt{\frac{m}{\pi T_{\pa}}} | \ln \delta| \frac{1}{v_{\perp}}\pder{v_{\perp}}\left(v_{\perp} \pder[f_{\perp}]{v_{\perp}} \right).
\end{equation}
Now, equation (\ref{perpendicular_equation}) becomes
\begin{equation}
-\frac{1}{2\tau_{r}v_{\perp}}\pder{v_{\perp}}(v_{\perp}^{2}f_{\perp}) = \sigma \sqrt{\frac{m}{\pi T_{\pa}}} | \ln \delta| \frac{1}{v_{\perp}}\pder{v_{\perp}}\left(v_{\perp} \pder[f_{\perp}]{v_{\perp}} \right),
\end{equation}
which can be solved for the perpendicular distribution function via direct integration to give
\begin{equation}
f_{\perp}(v_{\perp}) = \frac{m}{2\pi T_{\perp}} \exp \left( -\frac{m\vpp^{2}}{2T_{\pp}} \right).
\label{fperp}
\end{equation}
That is, in the scattering regime (II), $f_{\perp}$ is also a Maxwellian, but with a perpendicular temperature given through the equation 
\begin{equation}
\frac{T_{\perp}}{T_{\parallel}} = \frac{8}{\sqrt{2\pi}} \frac{\tau_{r}}{\tau_{2}} |\ln \delta |.
\label{624}
\end{equation}
That is, we see that the cooling during the scattering regime occurs on the timescale of parallel (to the magnetic field) collisions, and is thus a self-accelerating process as the temperature $T_{\parallel}$ falls. 

We can also see from equation (\ref{624}) that, in fact, $\delta$ and $\epsilon$ coincide within a factor of $A|\ln\epsilon|,$ where $A$ is some multiplicative $O(1)$ factor, and can therefore be treated as identical if we are willing to accept a relative error of order
$1/ | \log \epsilon |.$

Note, here equation (\ref{fperp}) is simply a specialisation of the result obtained in (I) when the distribution function can be factored into perpendicular and parallel components.

\section{Stability of anisotropic electron-positron plasmas in straight field-line geometry}\label{sec:7}

Having derived a kinetic theory for the radiative cooling of strongly anisotropic pair plasmas, we finish by modifying one of the important calculations of \cite{Helander2014}. We will show that stability of electron-positron plasmas to low frequency waves in slab geometry (i.e., a plasma threaded by a straight and constant magnetic field), still holds if the distribution function is non-Maxwellian. 

We write the gyrokinetic distribution function for our plasma as
\begin{equation}
f_{a}(\bm{r}) = F_{a} + e_{a} \phi (\bm{r}) \pder[F_{a}]{E} + \frac{e_{a}}{B_{0}}\left[\phi(\bm{r}) - \langle \phi(\bm{r})\rangle_{\bm{R}} \right]\pder[F_{a}]{\mu} + g_{a}(\bm{R})
\end{equation}
where $F_{a}$ is the apriori arbitrary equilibrium distribution function of species $a,$ $\phi$ is the electrostatic potential and $E$ is the energy of the plasma. We take care to distinguish between quantities which are evaluated at the particle position $\bm{r}$ and those which are evaluated at the guiding centre position $\bm{R}.$ The function $g_{a}$ satisfies the linearised, electrostatic, collisionless gyrokinetic equation \citep{CattoTang}:
\begin{equation}
iv_{\parallel}\gradd_{\parallel}g_{a} + (\omega - \omega_{da})g_{a} = -  e_{a} \bes \phi(\bm{R}) \left[ \omega \pder[F_{a}]{E} + \frac{1}{e_{a}B_{0}} \bm{k}\cross \bm{b} \cdot \gradd F_{a}   \right],
\end{equation}
where we have introduced $\omega_{da} = \bm{k}_{\perp} \cdot \bm{v}_{da}$ the magnetic drift frequency. We have also introduced the gyroaverage $\langle \alpha \rangle_{\bm{R}}$ of any function of the gyroangle $\alpha(\theta,\cdots)$ at fixed guiding centre $\bm{R},$ defined by 
\begin{equation}
\langle \alpha \rangle_{\bm{R}} = \left.\frac{1}{2\pi} \int_{0}^{2\pi} \alpha(\theta,\cdots) \right\vert_{\bm{R}=\text{const.}} \, \dd\theta.
\end{equation}

The quasineutrality condition demands
\begin{equation}
\sum_{a} e_{a} \int_{\bm{r} = \text{const.}} \dd^{3}\bm{v} \, f_{a} = 0,
\end{equation}
which for a pure electron-positron plasma requires 
\begin{equation}
\frac{1}{2e} \int_{\bm{R}=\text{const.}} \dd^{3} \bm{v} \, (g_{p} - g_{e}) J_{0} + \phi \int_{\bm{R}=\text{const.}} \dd^{3}\bm{v} \, \left( \mathrm{e}^{\eik} \pder[F]{E} + \frac{1}{B_{0}}\left[\mathrm{e}^{\eik} - J_{0} \right] \pder[F]{\mu} \right) J_{0}  = 0.
\end{equation}

In the limit of straight field lines we can set $\omega_{da} = 0$ and obtain the solution of the gyrokinetic equation as 
\begin{equation}
g_{a}(\bm{R}) = - \frac{e_{a} J_{0}\phi(\bm{R})}{\omega - k_{\parallel}v_{\parallel}} \left[ \omega  \pder[F_{a}]{E} + \frac{1}{e_{a}B_{0}} \bm{k}\cross \bm{b} \cdot \gradd F_{a}   \right],
\end{equation}
hence one obtains
\begin{equation}
g_{p} - g_{e} = -\frac{2\omega e J_{0} \phi}{\omega - k_{\parallel}v_{\parallel}} \pder[F]{E}.
\end{equation}
Substituting into Poisson's equation gives the dispersion relation
\begin{equation}
\int \dd^{3} \bm{v} \, \frac{\omega }{\omega - k_{\parallel}v_{\parallel}} \pder[F]{E} J_{0}^{2} = \int \dd^{3}\bm{v} \, \left( \mathrm{e}^{\eik} \pder[F]{E} + \frac{1}{B_{0}}\left[\mathrm{e}^{\eik} - J_{0} \right] \pder[F]{\mu} \right) J_{0}.
\end{equation}
One can see that, just as in \citep{Helander2014}, the gradients, which can drive microinstabilities, simply cancel due to the mass symmetry and the result that any instability must involve magnetic curvature still holds for anisotropic plasmas.

\section{Conclusions}\label{sec:8}
Radiative cooling will be an important process in upcoming pair plasma laboratory experiments and can lead to a strongly anisotropic distribution function. In the present contribution, we have explored the kinetic theory of plasmas including the effects of cyclotron emission. Specifically, we have investigated the evolution of the plasma kinetic energy in the collisionless regime, both in a straight field line geometry and also the extension of this result to the general case of a varying magnetic field. The influence of radiative cooling leads to the plasma dissipating its perpendicular energy on the radiation timescale $\tau_{r}$, initially the fastest dynamical timescale present in the system. 

Eventually, the collision time will become comparable to the radiation time and we therefore also investigated the experimentally relevant regime including both the long-range Coulomb collisions, which arise in strongly magnetized plasma, and also the standard short range collisions. The evolution of the plasma can then be split into three regimes, each of which is governed by different physics as set out in Section \ref{sec:5}. Taking advantage of the strong anisotropy allows the rate of change of energy to be found in each instance. In particular, it was found that the cooling during the scattering regime occurs on the timescale of parallel (to the magnetic field) collisions and is, thus, a self-accelerating process as the temperature $T_{\parallel}$ falls. 

The remarkable stability properties of electron-positron plasmas were found to persist in the strongly-anisotropic regime, at least insofar as the straight field-line case. The essence of this result is that the equilibrium distribution used in the more general form of the gyrokinetic equation is in fact arbitrary. The mass symmetry which leads to the precise cancellation of the gradient terms responsible for driving instability still holds. 

Crucially, we found at each stage of the investigation several results which could be useful in the successful operation of the first pair-plasma experiments. We found that the plasma still ought to be robustly stable to low-frequency microinstabilities and hence, potentially turbulence free, in spite of being strongly anisotropic. It was also found that the kinetic theory presented here leads to a reduction of the Debye length on the time scale of radiation emission or Coulomb collisions, whichever is slowest. If the particle confinement time exceeds both of these time scales this mechanism can be used to convert a collection of electron-positron pairs into the first electron-positron plasma. 

\vspace{1em}

We acknowledge Thomas Sunn Pedersen and the PAX/APEX experiment team for their interest in our work.

\newpage
\bibliographystyle{abbrvnat}
\bibliography{radiative_cooling_2.bib}

\end{document}